%% file: aa-v1.tex
\documentclass{aa}  

\usepackage{graphicx}
\usepackage{tabularx}
\usepackage{color}
\usepackage{txfonts}

\newcommand{\gc}{\bf{}}
\renewcommand{\gc}{} 

\begin{document} 
  
\title{A 80 au cavity in the disk around HD~34282}
	
\author{G.  van der Plas \inst{1,2}
        \and F.  M\'enard \inst{1}
        \and H.  Canovas \inst{3}
        \and H.  Avenhaus \inst{4,5}
        \and S.  Casassus \inst{2,5}
        \and C.  Pinte \inst{1}
        \and C.  Caceres \inst{6}
        \and L.  Cieza \inst{5,7}
}

   \institute{Univ.  Grenoble Alpes, CNRS, IPAG (UMR 5274), F-38000 Grenoble, France
        \and Departamento de Astronomia, Universidad de Chile, Casilla 36-D, Santiago, Chile
        \and Departamento de F\'isica Te\'orica, Universidad Aut\'onoma de Madrid, Cantoblanco, 28049 Madrid, Spain.
        \and ETH Zurich, Institute for Astronomy, Wolfgang-PauliStrasse 27, CH-8093, Zurich, Switzerland
        \and Millenium Nucleus Protoplanetary Disks in ALMA Early Science, Universidad de Chile, Casilla 36-D, Santiago, Chile
        \and Departamento de Ciencias Fisicas, Facultad de Ciencias Exactas, Universidad Andres Bello.  Av.  Fernandez Concha 700, Las Condes, Santiago, Chile
        \and Nucleo de Astronomia, Facultad de Ingenieria, Universidad Diego Portales, Av Ejercito 441, Santiago, Chile
        }

 
  \abstract
   {Large cavities in disks are important testing grounds for the mechanisms proposed to drive disk evolution and dispersion, such as dynamical clearing by planets and photo-evaporation.}
   {We aim to resolve the large cavity in the disk around HD~34282, such as has been predicted by previous studies modeling the spectral energy distribution}
   {Using ALMA band 7 observations we study HD~34282 with a spatial resolution of 0.10 $\times$ 0.17\arcsec at 345 GHz.}%
   {We resolve the disk around HD~34282 into a ring between 0.24 and 1.15\arcsec ~{\gc (78$^{+7}_{-11}$ and 374$^{+33}_{-54}$ au adopting a distance of 325$^{+29}_{-47}$ pc)}.  The emission in this ring shows azimuthal asymmetry centered at a radial distance of 0.46\arcsec and a position angle of 135 \degr and an azimuthal FWHM of 51\degr.  We detect CO emission both inside the disk cavity and as far out as 2.7 times the radial extent of the dust emission.}
   {Both the large disk cavity and the azimuthal structure in the disk around HD~34282 can be explained by the presence of a 50 M$_\mathrm{jup}$ brown dwarf companion at a separation of $\approx$ 0.1\arcsec.}


   \maketitle
%

\section{Introduction}

    Protoplanetary disks are the birth environments of planetary systems.  How these planets form is an ongoing topic of debate which is informed by an increasing number of disks that show various degrees of evolution and dispersal, such as opacity cavities (transitional disks), gaps \citep[pre-transitional disks, e.g.][]{2011ARA&A..49...67W}, and asymmetrical (lopsided) emission feautures.  Examples of such disks imaged at (sub) mm wavelengths include HD~100546 \citep{2014ApJ...791L...6W}, Sz~91 \citep{2015ApJ...805...21C, 2016MNRAS.458L..29C}, HD~142527 \citep{2013Natur.493..191C}, HD~97048 \citep{2017A&A...597A..32V} and SAO~206462 \citep{2009ApJ...704..496B}.  

    The common denominator between these disks is that their structure can be described by one large cavity or a broad ring of dust grains at reasonably large radii combined with multiple rings/gaps and/or azimuthal 'horseshoe-shaped' asymmetries in the outer disk.  The gaps and/or cavities in these disks are not empty: they contain both smaller dust grains, as traced by scattered light imaging \citep[e.g.  ][]{2012ApJ...745....5K, 2014ApJ...781...87A}, and gas, as most readily traced by rotational \citep{2015ApJ...798...85P, 2015A&A...579A.106V} and ro-vibrational carbon monoxide (CO) lines \citep{2009A&A...500.1137V, 2011ApJ...733...84P, 2014A&A...567A..51C, 2015A&A...574A..75V}.  Recently, long baseline Atacama Large Millimeter Array (ALMA) observations of HL Tau \citep{2015ApJ...808L...3A}, HD~163296 \citep{isella2016}, and TW Hya \citep{2016ApJ...820L..40A, 2016ApJ...819L...7N, 2016ApJ...829L..35T} have demonstrated that these disks show a rich substructure of many concentric rings and gaps at scales as small as 1 au when observed at very high spatial resolution.  It is indeed possible that most disks contain similar detailed structures that have not yet been resolved \citep{2016ApJ...818L..16Z}.
        
    Radial and azimuthal asymmetries in the dust emission structure of disks reflects variations in either the underlying mass, the grain properties (size distribution), and/or the temperature.  An often invoked mechanism to explain these structures is as response to local pressure maxima.  A local maximum in the gas density traps preferentially larger grains and thus stops the inward drift motion caused by the aerodynamic drag of the gas on the dust \citep{1977MNRAS.180...57W}, allowing dust grains to accumulate in these so-called dust-traps.  Such local gas pressure maxima have many proposed origins, for example being generated at the edge of a dead zone \citep{2010A&A...515A..70D}, by MRI instabilities \citep{2011ApJ...736...85U}, or at the edge of a planet-carved gap \citep[e.g.][]{2012A&A...545A..81P}.  They can even form spontaneously in simulations when the growth and fragmentation of dust grains and the back-reaction of the dust grains on the gas is taken into account \citep{2017MNRAS.467.1984G}.  Jumps in the pressure gradient can also induce vortices due to the Rossby wave instability \citep[RWI, e.g.][]{1999ApJ...513..805L, 2013ApJ...775...17L}.  Such vortices are especially promising places for planetesimal formation given their high efficiency in concentrating large dust masses \citep{2012A&A...545A.134M}.  A RWI however is only stable in low viscosity disks with $\alpha ~ \lesssim$ 10$^{-4}$\citep{2007A&A...471.1043D}.  Another mechanism proposed to explain horseshoe-shaped features in disks is the presence of an unequal-mass binary companion.  For large enough mass ratios the cavity becomes eccentric causing an azimuthally localized gas overdensity on the outer edge of the cavity \citep{2017MNRAS.464.1449R}.
         
    The detection and characterization of more structured disks is necessary to determine the physical origin of these asymmetries.  Disks around the intermediate mass Herbig Ae/Be (HAeBe) stars are good candidates for spatially resolving such structure, given their higher brightness and larger size compared to disks around the more abundant but lower mass/luminosity T Tauri stars.  These disks have historically been split into two groups based on the shape of their spectral energy distribution.  Group I sources display relatively bright  mid- to far-infrared emission and have been interpreted as hosting gas-rich protoplanetary disks with a flared, bright dust surface.  Most dust in group II disks is assumed to have settled towards the mid-plane and these disks therefore emit weaker mid- to far-infrared emission \citep{2001A&A...365..476M, 2004A&A...417..159D}.  Recent modeling of resolved observations of group I sources suggests that their bright infrared emission should instead be attributed to the large vertical walls on the limit of (large) dust cavities \citep{2012ApJ...752..143H,2013A&A...555A..64M}, an idea that is supported by high resolution scattered light imaging \citep{2017arXiv170301512G} and spatially resolved CO ro-vibrational observations \citep{2015A&A...574A..75V}.\\

    In this manuscript we present ALMA observations of the group I disk around the Herbig Ae/Be star  \object{HD~34282}.  The disk around this star has already been resolved using 1.3 mm continuum emission with a FWHM of 1.74\arcsec $\times$ 0.89\arcsec \citep{2003A&A...398..565P} as well as with rotational CO lines \citep{2000Icar..143..155G,2003A&A...398..565P}.  \citet{2009A&A...502L..17A} predicted the presence of an opacity cavity in this disk based on modelling the SED.  \citet{2016A&A...587A..62K} use spatially resolved Q band emission to estimate a gap-size 92 (+31, -17) au.
    
    Multiple estimates of the stellar parameters for HD~34282 exist.  We use the values derived by \citet{2004A&A...419..301M} that are based on detailed modeling of the stellar spectrum and the spectral energy distribution.  They find low metal abundances in the stellar spectrum and derive an A3 V spectral type, an age of 6.4$^{+1.9}_{-2.6}$ Myr, a luminosity of 13.64$^{+5.36}_{-12.02}$ L$_\odot$,  and a mass of 1.59$^{+0.30}_{-0.07}$ M$_\odot$ for the central star.  For source distance we use the value of 325$^{+29}_{-47}$ pc from the Gaia DR1 \citep{2016A&A...595A...1G, 2016A&A...595A...2G}.  This value is within the 1$\sigma$ values on the distance determined by \citet{2004A&A...419..301M} of 348$^{+129}_{-77}$ pc and of 400$^{+170}_{-100}$ pc \citep{2003A&A...398..565P}.  

\section{Observations and data reduction}

    ALMA Early Science Cycle 2 observations were conducted in the compact C43-2 configuration on December 12$^{\mathrm{th}}$ 2014 with 684 seconds of total time on source and in an extended C34-7 configuration on August 31$^{\mathrm{st}}$ 2015 with 342 seconds of total time on source.  The array configuration provided baselines ranging between respectively 15.1 and 348.5 meters, and between 15.1 and 1466.2 meters.  During the observations the precipitable water vapor had  a median value at zenith of respectively 0.811 and 0.755~mm.  

    Two of the four spectral windows of the ALMA correlator were configured in Time Division Mode (TDM) to maximise the sensitivity for continuum observations (128 channels over 1.875~GHz usable bandwidth).  These two TDM spectral windows were centred at 345.8~GHz and 356.7~GHz.  The other two spectral windows were configured in Frequency Division Mode (FDM) to target the  $^{12}$CO J=3-2 and the HCO+ J=4-3 lines with a spectral resolution of 105 and 103 m s$^{-1}$ respectively, using 0.23~GHz total bandwidth.  The data were calibrated and combined using the \textit{Common Astronomy Software Applications} pipeline \citep[CASA, ][version 4.5]{2007ASPC..376..127M}.  Inspection of the calibrated visibilities shows a 20\% difference in amplitude between the two observations at short baselines.  Given that the estimated error on the flux calibrator for the extended array configuration observations is twice as large as that for the compact configuration observations and assuming that the emission from the midplane is constant in the 8 month period spanning the observations, we scale the amplitudes of the visibilities for the extended array configuration observations to those of the compact array configuration observations.  Details of the observations and calibration are summarized in Table \ref{table_obs}, we estimate the absolute flux calibration to be accurate within $\sim$20\%.  

\begin{figure}
       \centering
        \includegraphics[width=\hsize]{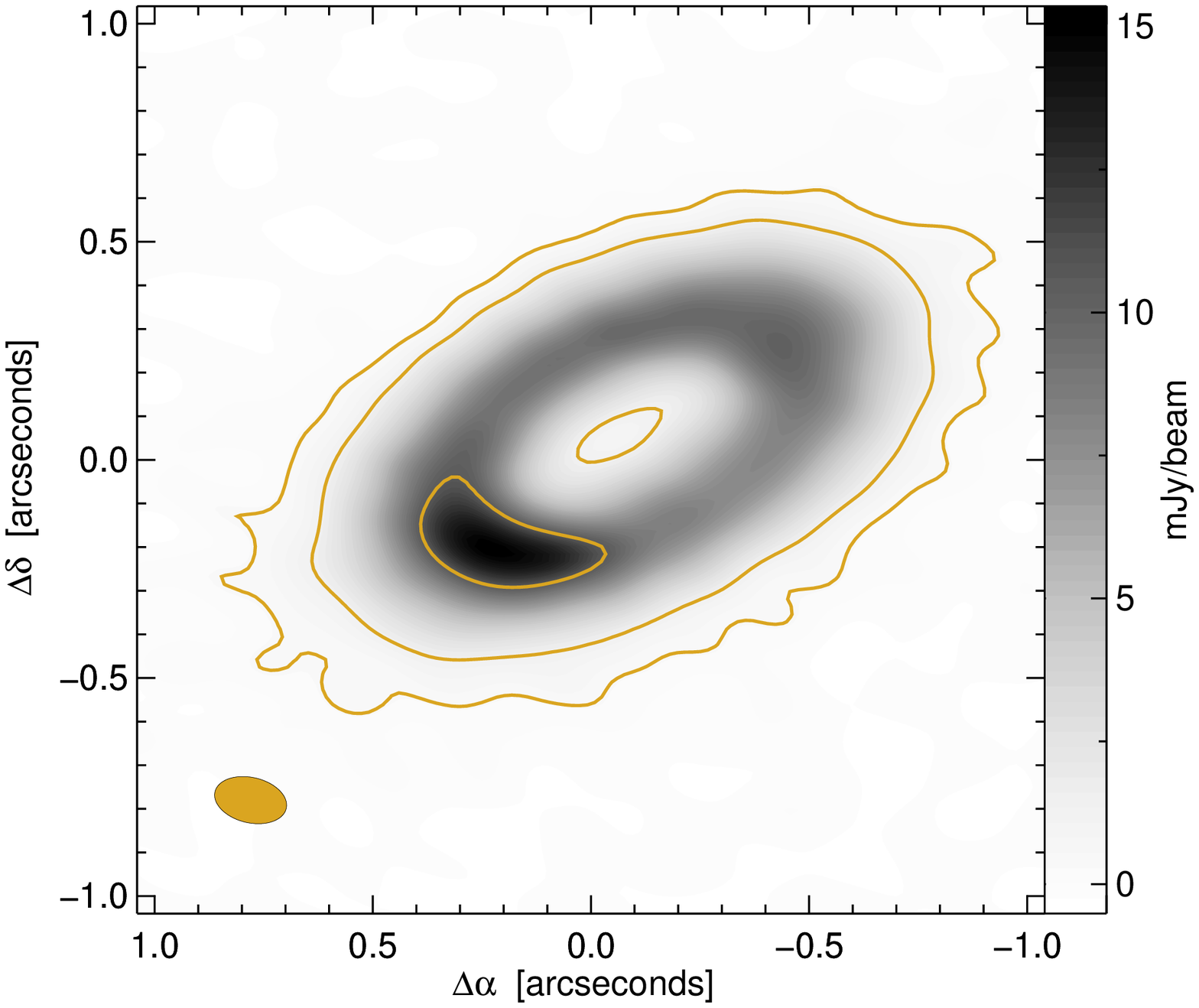}
        \caption{Continuum image of \object{HD~34282} for the ALMA band 7 (351.24 GHz or 0.853 mm) observations, reconstructed using superuniform weighting resulting in a 0.10 x 0.17" beam.  Over plotted are contours at 5, 15 and 100 times the RMS value of 0.11 mJy/beam.  The beam is shown in orange in the bottom left.}
         \label{fig:continuum}
    \end{figure}

\begin{figure*}
        \centering
        \includegraphics[width=\hsize]{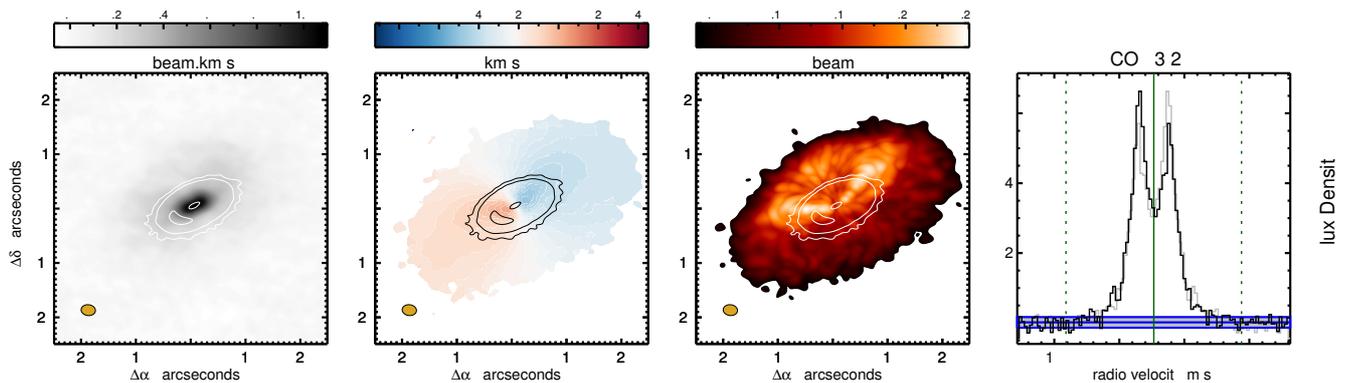}
        \caption{Summary of $^{12}$CO J=3-2 line emission from the disk around \object{HD 34282}.  We show the integrated intensity (moment 0, left panel), intensity-weighted velocity (moment 1, 2$^{\mathrm{nd}}$ panel), peak intensity (moment 8, 3$^{\mathrm{rd}}$ panel) and the collapsed emission line (right panel).  Each moment map was made using a 5 $\sigma$ cutoff and imaged using briggs weighting.  Over plotted in the three left panels are continuum contours with [5,15 and 100] times the RMS value of the continuum map imaged using superuniform weighting.  The beam is shown in orange in the bottom left of each panel.  The grey shaded area in the right panel denotes the + and - 3$\sigma$ level calculated outside the line boundaries.}
        \label{fig:moments-co32}
    \end{figure*}

\begin{figure*}
        \centering
        \includegraphics[width=\hsize]{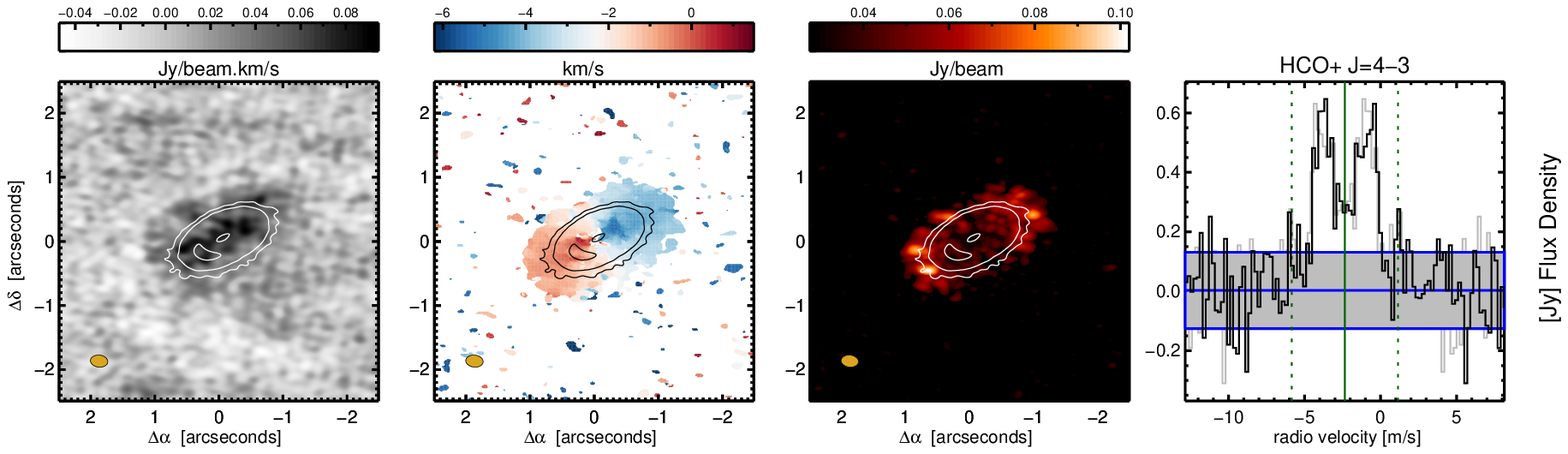}
        \caption{Summary of HCO+ J=4-3 line emission in \object{HD 34282}.  We show the integrated intensity (moment 0, left panel), intensity-weighted velocity (moment 1, 2$^{\mathrm{nd}}$ panel), peak intensity (moment 8, 3$^{\mathrm{rd}}$ panel) and the collapsed emission line (right panel).  Each moment map was made using a 3 $\sigma$ cutoff and imaged using briggs weighting.  Over plotted in the three left panels are continuum contours with [5,15 and 100] times the RMS value of the continuum emission.  The beam is shown in orange in the bottom left of each panel.  The grey shaded area in the right panel denotes the + and - 3$\sigma$ level calculated outside the line boundaries.}
        \label{fig:moments-hcop}
    \end{figure*}

    \input{table_obs}
    We image the disks with the CLEAN task in CASA \citep{1974AAS...15..417H} using briggs and superuniform weighting, which results in a restored beam of respectively  0.25" $\times$ 0.19" at PA = 88.3 degrees (briggs) and 0.17" $\times$ 0.10" at PA = 77.0 degrees (superuniform).  The dynamic range of these images is strongly limited by the bright continuum source and we perform self-calibration on both phase and amplitude, resulting in a final RMS of 0.11 mJy/beam for the images created using superuniform weighting.  We show the resulting continuum map in Figure \ref{fig:continuum}.  We applied the self-calibration solutions obtained from the continuum emission to the HCO+ and CO calibrated visibilities and subtracted the continuum emission using the CASA task \textit{uvcontsub}.  The resulting integrated intensity (moment 0), intensity-weighted mean velocity  (moment 1) and peak intensity (moment 8) maps and spectra are shown in Figures \ref{fig:moments-co32} and \ref{fig:moments-hcop} for the CO J=3-2 and HCO+ J=4-3 emission, respectively.

\section{Results}

    \subsection{continuum}\label{res:continuum}

        The HD~34282 dust continuum emission is concentrated into a ring that shows an azimuthal variation in intensity (Figure \ref{fig:continuum}).  We use the fitting library \textit{uvmultifit} \citep{2014A&A...563A.136M} to fit a superposition of simple geometrical shapes to the continuum visibilities in order to estimate the inclination, position angle and spatial distribution of the emission.  \textit{uvmultifit} minimizes the $\chi^2$ as function of the input model parameters.  Quoted parameter uncertainties are estimated from the post-fit covariance matrix which is scaled so that its $\chi^{2}_{r}$ has a value of 1.  
 
        \input{table_flux_uv}

        We fit the continuum visibilities to a combination of the following geometries: [1] a disk with a constant surface brightness, [2] an unresolved ring, and [3] a 2D Gaussian whose size and center are left free.  Each of these elements has the following free parameters: offset in RA and DEC from the phase center, flux, semi major axis, axis ratio, and position angle.  The disk element has two values for the semi major axis: one for the inner radius of the disk (the cavity radius), and one for the disk outer radius.  When fitting two disks or a disk and a ring simultaneously we force that they share the same center, inclination, and position angle.  We fit these shapes for each of the continuum windows (at 345.8~GHz and 356.7~GHz) separately to allow the detection of possible changes in flux due to the spectral slope of the dust emission $\alpha$ (S$_\nu$ $\propto$ $\nu^{\alpha}$).
                
\begin{figure*}
            \centering
            \includegraphics[width=\hsize]{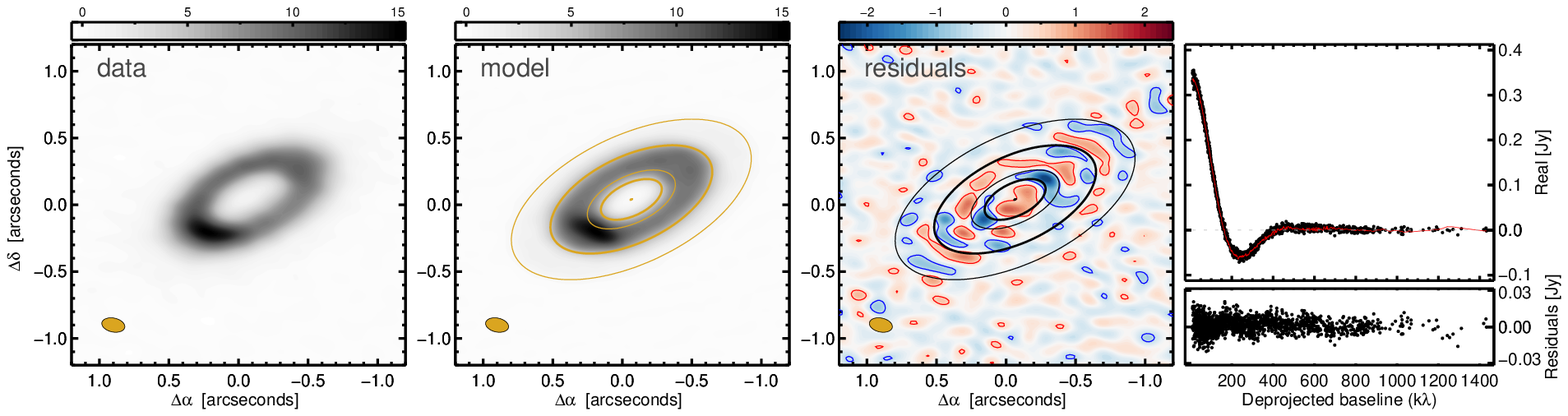}
            \caption{Comparison of ALMA band 7 data (left panel) with the best-fit composite 'disk + disk + gaussian' model (2$^{\mathrm{nd}}$ panel).  The 3$^{\mathrm{rd}}$ panel shows the imaged residuals.  We draw red and blue contours at 5 times the image rms of 0.11 mJy.  Units of all intensity scales are in mJy/beam.  The central two panels include ellipses representing the two fitted disk components to guide the eye.  These are drawn at the radial location corresponding to the inner and outer disk radius for disk \#1 (thick line) and disk \#2 (thin line), as summarized in Table \ref{tab:flux}.  The right panel (top) shows the real part of the visibilities as function of the deprojected baseline for the data (black dots) and the model (red line).  The bottom panel shows the residuals.  The visibilities are binned in sets of 200.}
            \label{fig:model_comparison}
        \end{figure*}

        We achieve the best fit with a combination of two disk components and a Gaussian that is offset from the disk center, as judged by minimizing the residuals of the calibrated visibilities of the data and our models.  The best fit parameters for the fitted components are listed in Table \ref{tab:flux} and are visualized in Figure \ref{fig:model_comparison}, where we compare the imaged model and residuals to the HD~34282 disk, together with the real part of the visibilities for the data, the model, and their difference.  Overall, our best fit provides a reasonable match to the data.  The center of the cavity is in agreement with the stellar position published in the Gaia DR1 \citep{2016A&A...595A...1G, 2016A&A...595A...2G}, and the dominant source of residuals is a radial pattern alternating between negative and positive signal most visible along the disk major axis.  This pattern is a consequence of using disk models with a radial constant surface brightness and a sharp drop at the disk boundary.  The most significant non-radially symmetric residuals are visible at the far side of the disk close to the disk minor axis, of the order of 1.0 mJy/beam, 9 times the image rms.

        We start our analysis with the following values derived from our best fit: a disk with a total flux of 323.9 mJy at 345.8 GHz (867 $\mu$m) with a spectral slope of -3.0 $\pm$ 0.7 coming from a disk that extends between 0.24 and 0.94\arcsec, an inclination of 59.3 degrees and a position angle of 117.1 degrees.  

        \subsubsection{The radial and azimuthal structure of the dust emission}\label{sec:cont_dep}

            From fitting the visibilities with a superposition of simple geometrical shapes we find that $\approx$ 95\% of the continuum emission is originating from a ring between 0.24 and 0.94\arcsec from the central star, while the remaining 5\% is concentrated in an region that is elongated in the azimuthal direction at a radial distance of $\approx$ 0.36\arcsec ~ from the star atop the smooth disk.  Looking at the imaged residuals (Fig.  \ref{fig:model_comparison}) there are hints of faint emission both inside the cavity and outside the fitted rings.
            
            The emission within the cavity, also visible in the imaged residuals, indicates that there is at least some dust inside the fitted cavity radius at 0.24\arcsec.  The amplitude of this residual emission is comparable to that of the residuals at the inner and outer boundaries of the fitted disks, and is likely caused by our choice of fitting a disk with a sharp edge.          
            \smallskip       
            
\begin{figure}
                \centering
                \includegraphics[width=\hsize]{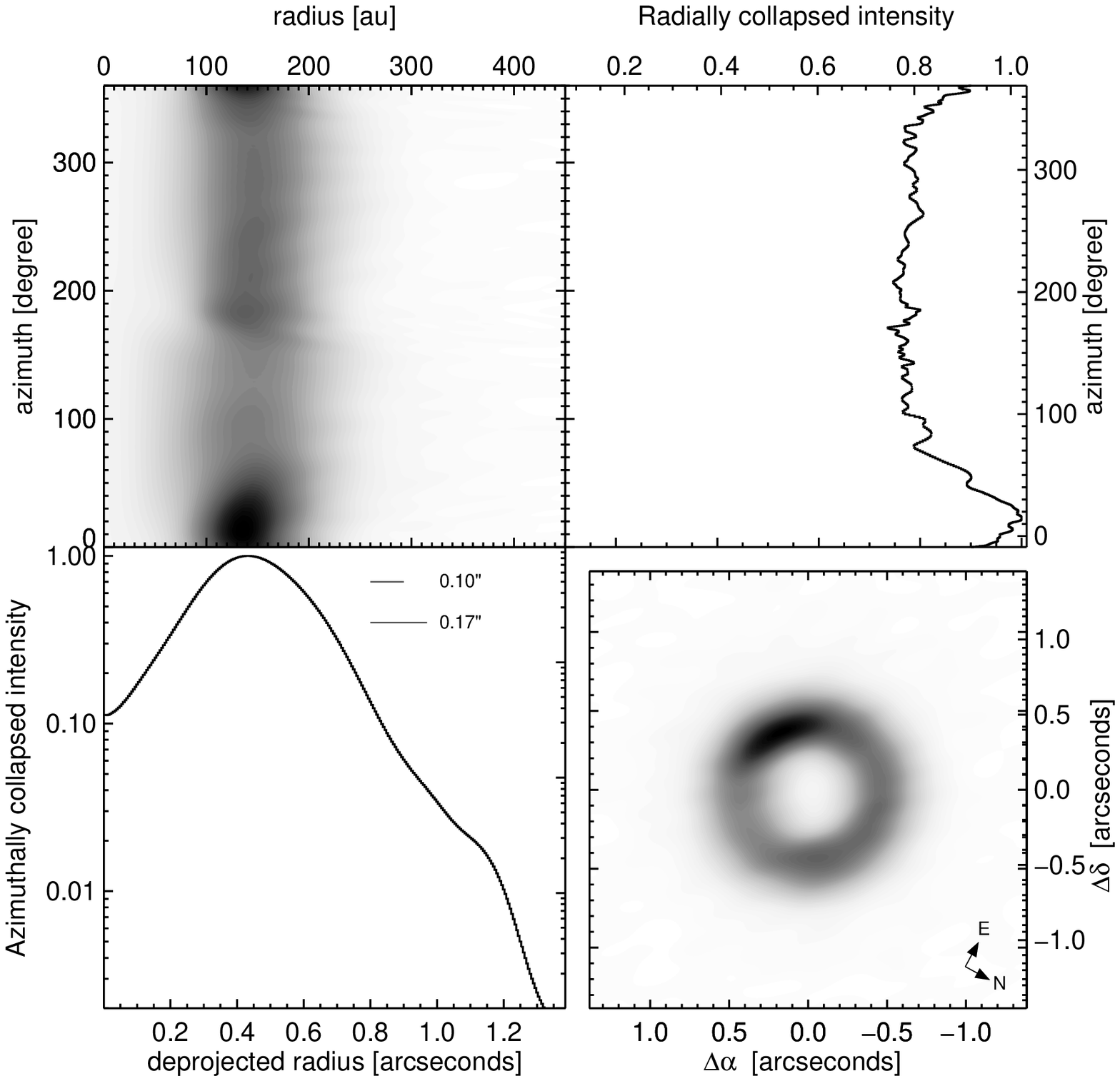}
                \caption{ALMA Band 7 image reconstructed using superuniform weighting, deprojected using the disk inclination and rotated with the position angle listed in Table \ref{tab:flux} to put the disk major axis at 0 degrees (north) (bottom right panel), converted to polar coordinates (top left panel).  This polar map is collapsed along the radial and azimuthal axes to yield the azimuthal intensity distribution (top right) and the radial intensity distribution (bottom left).  In this last panel we also show the size for the beam minor and major axis for reference.}
                \label{fig:radial_intensity}
            \end{figure}

            To better characterize the radial and azimuthal structure of the dust disk we remap the disk to polar coordinates after deprojecting it using the previously derived geometry (Figure \ref{fig:radial_intensity}).  The intensity profile in polar coordinates is shown in the top left panel, and is collapsed in the radial (top right panel) and azimuthal (bottom left panel) dimensions to obtain the respective integrated surface brightness structures.  
            
            The radial brightness profile for the continuum emission rises monotonically until it peaks at 0.44\arcsec.  After that it decreases and shows a change in slope around 1.15\arcsec until it becomes indistinguishable from the background around 1.35\arcsec, as shown in the bottom-left panel of Figure \ref{fig:radial_intensity}.
            The radial brightness distribution of the dust continuum emission between the peak at 0.44\arcsec and the outer disk radius at 1.35\arcsec can be fitted by two power-laws separated at 1.15\arcsec with exponents of -6 and -16, respectively.  
            
\begin{figure}
                \centering
                \includegraphics[width=\hsize]{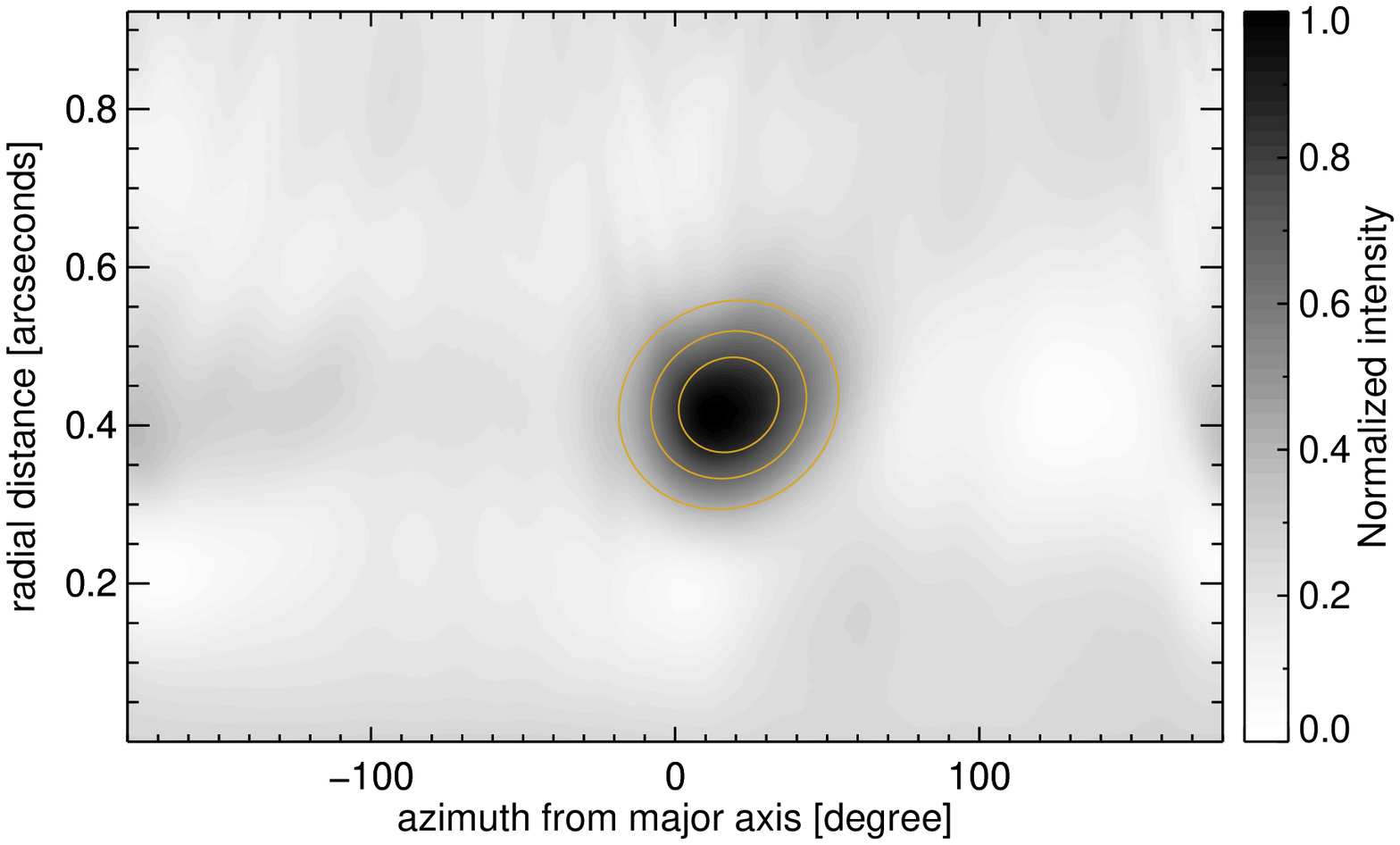}
                \caption{Surface brightness of the disk after subtracting a Gaussian fit to the median radial profile, shown in polar coordinates with the position angle relative to the disk major axis on the x axis and the radial distance from the center on the y axis.  Contours of the best fit 2D Gaussian are show at values of [0.25, 0.5 and 0.75] times its peak value.}
                \label{fig:dusttrap}
            \end{figure}

             The azimuthal brightness profile for the continuum emission peaks at an azimuth of 135\degr with a maximum $\approx$ 25\% above the median value for the disk and 18\degr away from the disk major axis.  To further study the azimuthal structure of the disk we subtract a median \textit{radial} Gaussian profile from each azimuthal row in the polar projection.  We create this average Gaussian by selecting only those radial slices with an integrated value within 1\% of the median integrated value over all azimuths (Fig.  \ref{fig:dusttrap}).  The remaining structure peaks close to the disk major axis and is best fit with a 2D Gaussian centered on (radius, azimuth) = (0.43\arcsec, 18\degr), with a FWHM of 0.15\arcsec in radius and 52\degr in azimuth.  This feature is discussed in \S \ref{sec:discussion}.  
	        
    \subsection{CO and HCO$^+$ emission} \label{res:gas}

        We detect spatially and spectrally resolved emission from the \element[ ][12]{CO} J=3-2 and \element[+][]{HCO} J=4-3 emission lines from the HD~34282 disk.  We show the moment maps and line profiles in Figures \ref{fig:moments-co32} and \ref{fig:moments-hcop} respectively.  
        
\begin{figure*}
            \centering
            \includegraphics[width=\hsize]{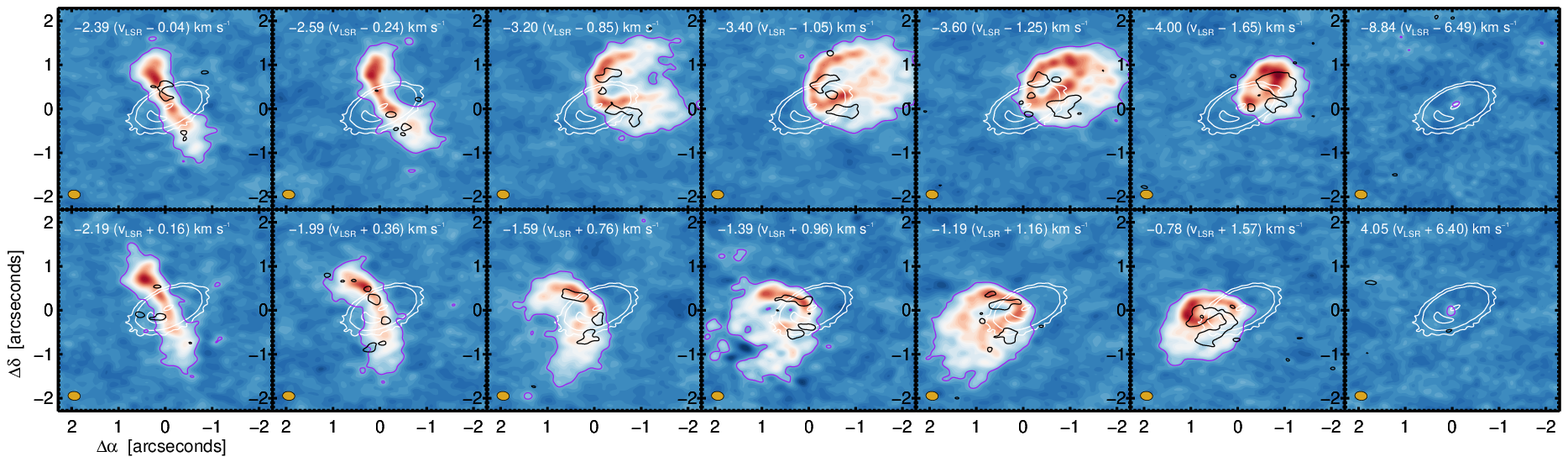}
            \caption{Selected channel maps showing the CO channels in color and its 3 $\sigma$ outline in purple, the HCO+ emission 3 $\sigma$ outline with black contours, and the continuum contours in white.  In the top right of each panel we note the v$_{lsr}$ in white noting the velocity with respect to the systemic velocity of -2.35 km s$^{-1}$ in parenthesis.  The clean beam is shown in yellow in the bottom left of each panel.} 
            \label{fig:channels}
        \end{figure*}

        We estimate the systemic velocity from the \element[ ][12]{CO} J=3-2 emission line at v$_{LSR}$ = -2.35 $\pm$ 0.10 km s$^{-1}$, based on the center of the line profile and the channel maps.  The line flux, integrated between + and - 6.75  km s$^{-1}$ from the systemic velocity is 21.2 $\pm$ 0.3 Jy km s$^{-1}$.  The semi major axis, as measured from the \element[ ][12]{CO} moment maps is 3.1\arcsec.  We detect \element[ ][12]{CO} emission coming from within the disk cavity and see two layers of CO emission on either side of the disk surface (Fig.  \ref{fig:channels}), which shows that the \element[ ][12]{CO} emission originates from the warm disk surface on both faces of the disk.  We calculate an upper limit on the inner radius of the gas disk assuming Keplerian rotation for the highest velocity at which we detect the line above 3$\sigma$, and measure the radial outer extent of the gas disk based on the presence of emission above 3$\times$ the RMS value.  We summarize the measured gas disk size and line flux in Table \ref{tab:flux_lines}.  In the naturally weighted channel maps CO emission $>$ 3$\sigma$ is present between -8.8 and +4.6 km s$^{-1}$.  These values can be translated to an emitting radius of $\approx$ 23 au, assuming the gas is in Keplerian rotation in a disk inclined by 59.3\degr around a 1.59 M$_{\odot}$ star.

        \input{table_lines}
        
        The low-J $^{12}$CO rotational emission lines in protoplanetary disks become optically thick quickly and trace a vertically thin region in the line of sight up to where the line becomes optically thick, making them a tracer of disk geometry \citep[e.g.][]{2013A&A...557A.133D, 2016A&A...586A.103W}.  This makes it possible to infer the disk geometry and spatial extend from the \element[ ][12]{CO} J=3-2 moment 8 (peak intensity) map shown in Figure \ref{fig:moments-co32}, panel 3.  Warmer CO gas emits stronger per unit volume, and the modulation of the peak brightness over the disk surface can be naturally interpreted as looking into a flared disk (bowl) where the far side of the disk is the warm and directly irradiated disk surface.  This warm surface is shielded by the flaring outer disk on the near (south-west) side, from which we instead see the cooler midplane and the backside of the disk.  

        The \element[+][]{HCO} emission is less extended with a semi major axis of 1.3\arcsec and is, with a flux of 2.2 $\pm$ 0.2 Jy km s$^{-1}$, 10 times weaker than the CO emission.  Based on the resolved emission in the channel maps the emission comes from deeper in the disk (closer to the midplane) and emission $>$ 3$\sigma$ is present between -5.8 and +1.2 km s$^{-1}$.  These values can be translated to an emitting inner radius of $\approx$ 85 au, assuming the gas is in Keplerian rotation in a disk inclined by 59.3\degr around a 1.59 M$_{\odot}$ star.  The \element[+][]{HCO} is brightest beyond the outer radius of the continuum ring, as seen in the moment 8 map in Figure \ref{fig:moments-hcop}.  We confirm that this is an artifact of subtracting the continuum from the \element[+][]{HCO} emission by repeating our analysis using the non-continuum-subtracted data.  This suggests that the continuum is absorbing part of the line emission mostly from the far side of the disk, or that the \element[+][]{HCO} emission is optically thick.\\
        
        onto the emitting CO surface through the midplane we see colder gas, which has lower emissivity.

\section{Discussion}\label{sec:discussion}

    In the following section we discuss the mass of the disk and its components, the spatial distribution of the dust and the gas, and possible mechanisms that can provoke such a system architecture.
    
    \subsection{Mass limits on the disk and the vortex-shaped feature}\label{sec:discussion_dust}

        A coarse but straightforward way to relate observed sub-mm fluxes to dust mass is by assuming the emitting dust is optically thin and of a single (average) temperature:
        
        \begin{equation}\label{eq:dustmass}
            log M_{dust} = log S_{\nu} + 2 log d - log \kappa_{\nu} - log B_{\nu}( \langle T_{dust} \rangle),
        \end{equation}
        
        where $S_{\nu}$ is the flux density, $d$ is the distance, $\kappa_{\nu}$ is the dust opacity, and  $B_{\nu}(\langle T_{dust} \rangle)$ is the Planck function evaluated at the average dust temperature {\gc \citep{1983QJRAS..24..267H}.  We adopt a dust opacity of 2.7 $cm^{2}~g^{-1}$ at 0.867 mm, calculated using smoothed UV astronomical silicate \citep{1984ApJ...285...89D,1993ApJ...402..441L, 2000AAS...197.4207W}, with a grain size distribution with sizes between 0.1 and 3000 $\mu$m distributed following a power law with a slope of -3.5. We } estimate the average dust temperature using the correlation with stellar luminosity $\langle T_{dust} \rangle \approx 25 (L_{*} /L_{\odot})^{1/4} K$ \citep[e.g.  ][]{andrews2013}, leading to a $\langle T_{dust} \rangle$ of 48 K.  This is most likely an overestimate of the dust temperature since all dust in the disk around HD~34282 is located outside $\approx$ 78 au, and the resulting dust mass should thus be interpreted as a lower limit.  Using these assumptions we calculate a dust mass of 0.41 M$_{\mathrm{jup}}$ for the disk, and 6.6 M$_{\mathrm{earth}}$ for the vortex-shaped feature.  With a dust-to-gas ratio of 1:100 this leads to a disk that weights 2.5\% of the stellar mass.  The line-of-sight optical depth calculated using the parameters of our best-fit disk model and using a dust opacity of 2.7 $cm^{2}~g^{-1}$ and an albedo of 0.75, is 1.1 in the vortex-shaped feature and 0.6 along the ring.

    \subsection{The radial and azimuthal structure of the dust emission}\label{sec:discussion_dust}

        Both defining features of the dust continuum emission, the ring and the azimuthal asymmetry, can be explained by the dynamics of gas and dust at the edge of the cavity if there is an unseen massive companion present.  \citet{2017MNRAS.464.1449R} use 3D SPH gas and dust simulations to test the effect of a binary pair of unequal mass on a circum-binary disk.  In their simulations the companion carves a wide and eccentric cavity resulting in a non-axisymmetric gas overdensity at the cavity edge.  The amount of azimuthal asymmetry in their simulations scales with the binary mass ratio: a ratio of 0.01 results in a ring-like structure, while a mass ratio of 0.05 produces a contrast ratio of $\approx$ 1.5 between the feature and the background disk.  Since the brightness contrast ratio in the HD~34282 disk is $\approx$ 1.25, a naive interpolation would put the companion mass at $\approx$ 0.05 M$_\odot$ (50 M$_{\mathrm{jup}}$).

	    The presence of a binary companion has already been suggested by \citet{2010MNRAS.401.1199W} based on the spectro-astrometric signal of the H$\beta$ line.  No estimate of the separation or position angle is given though, due to artifacts in the astrometric signal.  The spectro-astrometric technique used is sensitive to binaries with a separation between  0.1 and 2\arcsec with a brightness contrast of up to 5 magnitudes \citep{2006MNRAS.367..737B}, putting the companion somewhere between 0.1\arcsec and the cavity radius at 0.24\arcsec.  A separation close to 0.1\arcsec is in agreement with predictions by \citet{1994ApJ...421..651A} for the location of the cavity outer edge, who predict the inner edge for the circumbinary disk to be between 1.8 to 2.6 times the binary semi major axis for low (between 0 and 0.25) eccentricities.

        This scenario, where a brown dwarf binary companion induces the outer disk morphology, is backed up by the following arguments: 
        
        (1) Besides dynamical clearing, inside-out photoevaporation of the disk by the stellar radiation is often invoked as cause for disk cavities \citep[e.g.][]{2001MNRAS.328..485C, 2006MNRAS.369..216A,2010MNRAS.401.1415O}.  However, in the case of this disk we deem photo evaporation unlikely to have shaped the cavity.  The non-detection of CO gas close to the star in both the rotational (this work) and ro-vibrational lines \citep{2005A&A...436..977C} is consistent with a photo evaporation scenario.  However, an important caveat is the sensitivity of the latter CO observations.  Typical line strengths for CO rovibrational lines originating from disks around HAeBe disks are about 5 to 10\% above the continuum level \citep[e.g. ][]{2015A&A...574A..75V}.  Such lines could easily be  hidden in the noise given the spectrum presented in Figure 1 of \citet{2005A&A...436..977C}, which makes HD~34282 a prime target for high sensitivity observations of the fundamental ro-vibrational CO lines.   
        
        The arguments against photoevaporation as cause for the cavity are stronger.  Cavities carved out by radiation pressure from the central star are predicted to have a sharp edge, which is not the case for the HD~34282 disk.  The inside-out nature of photoevaporation as root-cause for the cavity also precludes the presence of any inner disk, while a halo or small inner disk is required to fit the NIR excess observed for this source \citep{2016A&A...587A..62K}.

        (2) The stellar Fe/H abundance for HD~34282 is strongly depleted, in line with the suggestion by \citet{2015A&A...582L..10K} that depletion of heavy elements emerges as companions block the accretion of part of the dust, while gas continues to flow towards the central star.\\

        The spectral index value we derive for the complete disk (-3.0 $\pm$ 0.7) is lower than, but within uncertainty consistent with, the canonical value of $\approx$ -2.3 for disks \citep{2014prpl.conf..339T}.  This lower value is in line with the trend observed in for example AS 209 where the grain size distribution is weighted more towards larger grains in the inner disk leading to a value of $\alpha$ $>$ -2.5 at 20 au, and increasing to $<$ -3.5 outside 80 au \citep{2012ApJ...760L..17P}.
        
        One out of the three components, the Gaussian offset from the disk center, in our best-fit disk model deviates  $>$ 1$\sigma$ from the typical value for young disks of -2.3, whereas the small value of $\alpha$ in the fitted Gaussian suggests advanced grain growth.  However, given the large error on this value we do not consider this deviation significant.  Upcoming work on this source comparing these data with 1.3~mm ALMA measurements at similar angular resolution will be better able to decide on the nature of the azimuthal brightness modulation.  This azimuthally asymmetric feature in the disk originates from the same radial distance as the peak emission of the dust ring as can be seen in Figure \ref{fig:dusttrap}.  Its radial size is comparable to the beam and thus could be unresolved.  The feature is resolved in the azimuthal direction with a FWHM of 52\degr.

    \subsection{Distribution and kinematics of the gas} 
    
        It is possible that CO gas is present closer to the star than the 24 au limit measured from our data as beam dilution would render the emission undetectable given the size of our beam (65 $\times$ 85 au for the briggs-weighted maps).  In the outer disk regions the CO gas can be traced out as far as 3.1\arcsec (1000 au), a factor of $\approx$ 2.7 further out than the dust detection coming from larger grains.  
    
\begin{figure}
        		\centering
        		\includegraphics[width=\hsize]{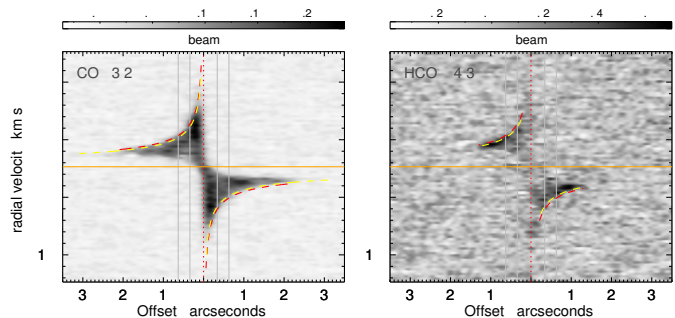}
        		\caption{Position-velocity diagram for the CO J=3-2 (left) and HCO$^+$ J=3-2 (right) emission made using the disk center and position angle listed in Table \ref{tab:flux}.  We use a dotted vertical red line to show the stellar position.  The systemic velocity of -2.75 km s$^{-1}$ is shown with an orange horizontal line.  We show the Keplerian rotation curves for the two different estimates for stellar mass and distance from \citet{2004A&A...419..301M} (yellow: M$_*$ = 1.59 M$_\odot$ and d = 348 pc) and from \citet{2003A&A...398..565P} (red: M$_*$ = 2.1 M$_\odot$ and d = 400 pc).  The radial inner and outer extent of the curves is taken from the inner and outer radius of the emission listed in Table \ref{tab:flux_lines}.  We also overplot in both panels with grey vertical lines the radial distances of 0.34\arcsec and 0.63\arcsec to mark the band where the continuum emission is strongest.}
       		 \label{fig:pv}
    	\end{figure}

        Judging by the velocity field, the CO emission is in agreement with that from a Keplerian rotating disk.  We verify our choice for distance (325 au) and stellar mass (1.59 M$_\odot$) by using the kinematics of the CO emission.  We construct a position-velocity map using the disk position derived from fitting the continuum emission and compare this with the expected Keplerian rotation curves for the stellar distance and mass derived by \citet{2003A&A...398..565P} and \citet{2004A&A...419..301M} respectively in Figure \ref{fig:pv}.  The inner radius for the CO rotation curves is set at 24 au, the outer radius either at 835 au \citep[from][]{2003A&A...398..565P}, or at 1080 au (The CO outer disk size derived in this work scaled to the stellar distance of \citet{2004A&A...419..301M}.  For the HCO+ PV diagrams we draw rotation curves between 85 au and an outer radius derived in this paper scaled the stellar distance (400 and 348 pc) in both papers.  
	
	    These PV diagram helps illustrate two points.  First, that our choice for the closer distance of 325 au together with a lighter stellar mass is still in agreement with the CO rotation curves.  Second, that the CO emission is suppressed at the location where the continuum emission is strongest, between 0.34 and 0.63\arcsec.  This latter is an artifact of the data reduction, where subtracting the continuum emission suppresses line emission at those locations where the emission is optically thick, or, in case of optically thin emission, where the dust in the midplane absorbs some emission coming from the far side of the disk.  
        
         We do not detect any kinematical deviation from Keplerian rotation in the CO emission within the gap as predicted for the interaction of Jupiter-mass planets with the gas disk \citep{2015ApJ...811L...5P}.  The observations we present are of the highly optically thick \element[ ][12]{CO} J=3-2 transition, and were made with both lower spatial and sensitivity compared to those predictions presented in \citet{2015ApJ...811L...5P}.  These 3 factors all conspire to make a possible kinematical imprint in the gas by the companion more difficult to detect.  A deeper observation on an optically thinner isotopologue such as \element[][13]{CO} at ALMA's highest resolution could help uncover a putative companion in HD~34282.

        The \element[+][]{HCO} J=4-3 emission is detected in closer proximity to the dust ring starting at the cavity outer radius and extending out to 1.3\arcsec.  Compared to the CO emission it is originating closer to the midplane, very similar to the emission detected from the HD~97048 disk \citep{2017A&A...597A..32V}.  Because \element[+][]{HCO} ions quickly disappear without gas-phase CO molecules \citep[e.g.][]{2014ApJ...794..123C} we interpret this height difference as a vertical temperature gradient.\\
        
    \subsection{Comparison of the radial extent of the dust and gas disk} 
    
        The radial extent of the disk as measured with \element[ ][12]{CO} J=3-2 emission is a factor of 2.7 larger than the disk measured at 345 GHz.  While such a difference can be largely explained by different optical depths for the dust and gas \citep[e.g.][]{1998A&A...338L..63D, 1998A&A...339..467G} without the need for radial drift of larger particles \citep{2017arXiv170506235F}, the sharp drop at 1.15\arcsec is very reminiscent of the drop seen in the e.g.  the disks around TW Hya \citep{2016A&A...586A..99H, 2016ApJ...820L..40A} and HD~97048 \citep{2017A&A...597A..32V}.  We follow the interpretation in those works that the sharp drop is caused by the outer edge of a drift-dominated dust distribution \citep{2014ApJ...780..153B, 2017arXiv170506235F} and the tail end of the radial intensity distribution as a smearing effect by the beam of the outer disk edge at 1.15\arcsec.

\section{Conclusions}

    We resolve the disk around HD~34282 in the dust continuum emission at 867 $\mu$m into a ring between 0.24\arcsec and 1.15\arcsec {\gc , or between 78$^{+7}_{-11}$~au and  374$^{+33}_{-54}$~au using a distance of 374$^{+33}_{-54}$}.  There is an azimuthal asymmetry present in the dust continuum emission that coincides with the radial position of the ring but is radially unresolved, with an azimuthal extent of 52\degr, and contains 5\% of the total sub-mm flux on top of the background disk emission.  We also detect The \element[ ][12]{CO} J=3-2 and \element[+][]{HCO} J=4-3 emission lines.  Assuming Keplerian rotation we detect CO emission between 24 au and 1000 au, 2.7 times as far out as the mm dust grains.  The sharp outer edge of the dust disk suggests that this is due to radial drift.\\
    
    We discount photo-evaporation as opening mechanism for the disk cavity.  Rather, the disk cavity and azimuthal structure, the presence of gas within the cavity, and the low stellar accretion, all can be explained by the presence of a $\approx$ 50 M$_{\mathrm{jup}}$ brown dwarf companion in the gap at a distance of $\approx$ 0.1\arcsec.

\begin{acknowledgements}
    This paper makes use of data from ALMA programme 2013.1.00658.S.  ALMA is a partnership of ESO (representing its member states), NSF (USA) and NINS (Japan), together with NRC (Canada) and NSC and ASIAA (Taiwan), in cooperation with the Republic of Chile.  The Joint ALMA Observatory is operated by ESO, AUI/NRAO and NAOJ.  The National Radio Astronomy Observatory is a facility of the National Science Foundation operated under cooperative agreement by Associated Universities, Inc.  This work has made use of data from the European Space Agency (ESA) mission {\it Gaia} (\url{http://www.cosmos.esa.int/gaia}), processed by the {\it Gaia} Data Processing and Analysis Consortium (DPAC, \url{http://www.cosmos.esa.int/web/gaia/dpac/consortium}).  Funding for the DPAC has been provided by national institutions, in particular the institutions participating in the {\it Gaia} Multilateral Agreement.  Gvdp, CC and SC acknowledge support from the Millennium Science Initiative (Chilean Ministry of Economy) through grant RC130007.  GP acknowledges financial support from FONDECYT, grant 3140393, LC acknowledges support from FONDECYT grant 1171246, and SC acknowledges support from FONDECYT grant 1130949.  GP and FM acknowledge funding from ANR of France under contract number ANR-16-CE31-0013 (Planet-Forming-Disks).  CC acknowledges support from CONICYT PAI/Concurso nacional de insercion en la academia 2015, Folio 79150049.  H.A.  acknowledges the financial support of the Swiss National Science Foundation within the framework of the National Centre for Competence in Research PlanetS and H.C.  acknowledges support from the Spanish grant AYA 2014-55840-P.
\end{acknowledgements}




\end{document}

%% file: table_obs.tex
\begin{table*}
\caption{Details of the observations.\label{table_obs}} \smallskip
\begin{minipage}[t]{\textwidth}
\centering
\noindent\begin{tabularx}{\columnwidth}{@{\extracolsep{\stretch{1}}}*{7}{l}@{}}
\hline\hline
UT Date & Number & Baseline Range & pwv &  \multicolumn{3}{c}{Calibrators:} \\ 
 & Antennas & (m) & (mm) & Flux & Bandpass & Gain  \\
\hline
2014 Dec 12	& 37 & 15.1 to 348.5  & 0.811 &  J0423-013 &  J0522-3627 & J0501-0159\\ 
2015 Aug 31	& 38 & 15.1 to 1466.2 & 0.755 &  J0423-013 &  J0423-0120 & J0542-0913 \\
\hline
\end{tabularx}
\end{minipage}
\end{table*}

%% file: table_flux_uv.tex
\begin{table*}
\caption{Best-fit parameters with their respective 1 $\sigma$ uncertainty in parenthesis, obtained from fitting separate components to the continuum visibilities: either one component (radially constant disk, a ring, and a Gaussian), or a combination of these components. When fitting a disk and ring component simultaneously, the following parameters were fixed between the two: The offset from the pointing center, the inclination and position angle. In the Table we represent these values as "fixed". The spectral slope $\alpha$ (5$^{\mathrm{th}}$ column) is calculated following S$_\nu$ $\propto$ $\nu^{\alpha}$. \label{tab:flux}} 
\begin{minipage}[t]{\textwidth}
\centering
\noindent\begin{tabularx}{\columnwidth}{@{\extracolsep{\stretch{1}}}*{9}{l}@{}}
\hline\hline
Component & $\Delta$RA$^{\mathrm{a}}$ & $\Delta$DEC$^{\mathrm{a}}$ & S$_{v, 345.8 GHz}$ & $\alpha$ & semi major axis & inclination & PA     &  X$^2_r$\\
   & [\arcsec]    & [\arcsec]     & [mJy]  &  & [\arcsec]         & [\degr]       & [\degr]   &       \\

\hline
1 component: & & & & & & & &  \\
\hline
Gaussian (\textbf{G}) & -0.039 & 0.017 & 338.0 (0.7) & -3.2 (0.1) & 0.53 (0.01) & 60.4 (0.4) & 119.4 (0.4) & 9.69 \\
Ring (\textbf{R}) & -0.046 & 0.022 & 312.3 (0.5) & -2.9 (0.1) & 0.50 (0.01) & 59.0 (0.3) & 116.7 (0.4) & 9.30 \\
Disk (\textbf{D}) & -0.044 & 0.021 & 322.4 (0.6) & -3.1 (0.1) & 0.19 (0.07), 0.74 (0.03)$^{\mathrm{b}}$ & 59.4 (0.6) & 117.5 (0.4) & 8.97 \\
\hline
$>$1 components: & & & & & & & &  \\
\hline
\textbf{D + R}&  & & & & & & &  \\
Disk 1  & -0.053 & 0.029 & 252.6 (1.3) & -3.2 (0.2) & 0.21 (0.02), 0.75 (0.01)$^{\mathrm{b}}$ & 59.7 (0.4) & 118.0 (0.1) & \\
Ring 1  & fixed & fixed  & 65.6 (1.3) & -1.6 (0.9) & 0.45 (0.01)  & fixed     & fixed      & \\
total &  &  &  318.2 (1.8) & -2.9 (0.4) &   &      &       & 5.76\\
\hline
\textbf{D + D}&  & & & & & & &  \\
Disk 1  & -0.053 & 0.029 & 215.1 (2.1) & -3.1 (0.5) & 0.26 (0.02), 0.61 (0.01)$^{\mathrm{b}}$ & 59.6 (0.4) & 117.9 (0.3) & \\
Disk 2  & fixed & fixed  & 108.1 (2.2) & -2.8 (1.0) & 0.32 (0.06), 0.90 (0.01)  & fixed     & fixed      & \\
total &  &  &  323.1 (3.0) & -3.0 (0.6) &   &      &       & 5.50\\
\hline
\textbf{D + D + G}  & & & & & & & & \\
Disk 1 & -0.063 & 0.039 & 221.2 (2.1) & -4.5 (1.1) & 0.24 (0.01), 0.63 (0.01)$^{\mathrm{b}}$ & 59.3 (0.4) & 117.1 (0.3) &  \\
Disk 2 & fixed & fixed  & 87.0  (2.1)  & -2.5 (0.5) & 0.34 (0.07), 0.94 (0.01)      & fixed     & fixed      &  \\
Gaussian & 0.270$^{\mathrm{c}}$ & -0.240$^{\mathrm{c}}$ & 15.7 (0.3) & -1.5 (1.1) & 0.15 (0.01)     & 70.1 (1.0) & 67.4 (4.6)  & \\
total &  &  &  323.9 (3.0) & -3.0 (0.7) &   &      &       & 5.41\\
\hline

\end{tabularx}
\end{minipage}
\tablefoot{\textbf{$^{\mathrm{a}}$:} Offset from the pointing center. \textbf{$^{\mathrm{b}}$:} Contains two values for the disk component: the inner and outer radius. \textbf{$^{\mathrm{c}}$:} Offset relative to the center of the best-fit disk and ring component.}
\end{table*}

%% file: table_lines.tex
\begin{table}
\setlength{\tabcolsep}{4pt} 

\caption{Line fluxes, spectral resolution and spatial extent for the CO J=3-2 and HCO$^+$ J = 4-3 lines\label{tab:flux_lines}} 
\smallskip
\centering   
\scriptsize
\noindent\begin{tabularx}{\columnwidth}{@{\extracolsep{\stretch{1}}}*{7}{l}@{}}
\hline

Line  & line flux & error$^{\mathrm{a}}$ & channel width          & RMS$^{\mathrm{b}}$   & major axis$^{\mathrm{c}}$ & r$_{\mathrm{in}}^{\mathrm{d}}$\\
      & Jy  km s$^{-1}$  &  Jy   km s$^{-1}$ & m s$^{-1}$  & mJy/beam          &  \arcsec  & au \\
\hline


CO J = 3-2      &  21.2 & 0.3 & 200 & 9.4 & 6.2 & 23 \\
HCO$^+$ J = 4-3 &  2.2  & 0.2 & 200 & 10.4 & 2.6 & 85 \\

\hline                                   
\end{tabularx}
\tablefoot{Line fluxes have been calculated from the briggs-weighted images by integrating the emission around the  systemic velocity at -2.35 km s$^{-1}$ assuming a half line width of 3.5 km s$^{-1}$ for the HCO$^+$ J = 4-3 line and 6.75 km s$^{-1}$ for the CO J = 3-2 line. \textbf{$^{\mathrm{a}}$:} Estimated from the RMS of the integrated spectrum outside the line boundaries, does not include calibration uncertainties. \textbf{$^{\mathrm{b}}$:} 1 $\sigma$ RMS per channel. \textbf{$^{\mathrm{c}}$:} Major axis is determined for all emission above 3 times the RMS per channel. \textbf{$^{\mathrm{d}}$:} Calculated from the maximum velocity for which emission $>$ 3$\sigma$ is present in the channel maps, assuming the gas is in Keplerian rotation in a disk inclined with 59.3\degr around a 1.59 M$_\odot$ star.}
\end{table}